\documentclass[12pt,a4paper]{article}
\usepackage[inner=2.5cm,outer=2.5cm,top=1cm,bottom=2cm,includeheadfoot]{geometry}				%Einstellungen der Seitenränder

\usepackage{amsmath}
\usepackage{amssymb}
\usepackage{mathrsfs}  
\usepackage{braket}
\usepackage{cite}

\usepackage[affil-it]{authblk}
\usepackage[utf8]{inputenc}
\usepackage{url}
\usepackage{hyperref}
\usepackage{graphicx}
\usepackage[dvipsnames]{xcolor}
\usepackage[T1]{fontenc}
\usepackage{booktabs}
\usepackage{multirow}
\usepackage{rotating} 
\newcommand{\del}{\partial}

\newcommand{\e}{\mathrm e}
\renewcommand{\d}{\mathrm d}
\newcommand{\mi}{\mathrm i}
\renewcommand{\(}{\left(}
\renewcommand{\)}{\right)}
\renewcommand{\[}{\left[}
\renewcommand{\]}{\right]}

\newcommand{\NLSM}{{{\rm NL}\sigma{\rm M}}}

\begin{document}

\title{Precise test of Higgs properties via
triple Higgs production  in VBF at future colliders}
\author[a,b]{A. S. Belyaev\footnote{Email:A.Belyaev@soton.ac.uk}}
\author[a]{P. B. Schaefers\footnote{Email:P.Schaefers@soton.ac.uk}}
\author[a]{and M. C. Thomas\footnote{Email:M.C.Thomas@soton.ac.uk}}

\affil[a]{School of Physics \& Astronomy, University of Southampton,\\ Southampton SO17 1BJ, UK}
\affil[b]{Particle Physics Department, Rutherford Appleton Laboratory,\\ Chilton, Didcot, Oxon OX11 0QX, UK}

% e-mail addresses: one for each author, in the same order as the authors
%\ead{A.Belyaev@soton.ac.uk}
%\ead{P.Schaefers@soton.ac.uk}
%\ead{M.C.Thomas@soton.ac.uk}

\maketitle

\begin{abstract}
\noindent
For certain classes of Beyond the Standard Model theories, including composite Higgs models,
the coupling of the Higgs to gauge bosons can be different from the Standard Model one.
In this case, the multi-boson production via vector boson fusion (VBF)
can be hugely enhanced in comparison to the SM production one due to the lack of cancellation 
in longitudinal vector boson scattering.
Among these processes, triple Higgs boson production in VBF plays a special role ---
its enhancement is especially spectacular due to the absence of background from transversely polarised 
vector bosons in the final state.
While the rates from $pp\to jjhhh$ production in vector boson fusion are too low at the LHC and even at 
future 33 TeV $pp$ colliders, we have found that the 100 TeV $pp$ future circular collider (FCC) has the unique opportunity to probe the $hVV$ coupling far beyond the LHC sensitivity.
We have evaluated the $pp\to jjhhh$ rates as a function of deviation from the $hVV$ coupling and 
have found that the background is much smaller than the signal for
observable signal rates. We also found that the 100 TeV $pp$ FCC 
can probe the $hVV$ coupling up to the permille level, which is far beyond the LHC reach.
These results highlight a special role of the $hhh$ VBF production and stress once more
the importance of the 100 TeV $pp$ FCC.
\end{abstract}

%\keywords{}

\newpage
\tableofcontents
%\newpage

\section{Introduction}
\label{sec:introduction}
In 2012, a new scalar particle was discovered during the first run of the LHC
with a collision energy of $\sqrt{s} = 8$ TeV
\cite{Aad:2012tfa,Chatrchyan:2012ufa}. Although the found particle is thought to
fit the Standard Model (SM) Higgs boson astonishingly well, it is still possible
that it belongs to a different theory such as a composite Higgs model,
Supersymmetry or some other theory.

The increase of LHC energy and luminosity
as happened in LHC Run 2 has allowed  to  understand Higgs boson properties more precisely.
However, this increase in not sufficient to measure Higgs boson properties at the 
percent level or below.
For this purpose, future colliders with collision energies up to 33 TeV (LHC)
or 100 TeV ($pp$ future circular collider
(FCC)) are being planned \cite{Ball:2014abc}. While
not built yet, there already is a broad range of prospects and predictions for
various topics  including Higgs physics and
supersymmetry (see e.g.
\cite{He:2014xla,Barr:2014sga,Low:2014cba,Acharya:2014pua,Auerbach:2014xua,Fowlie:2014awa,Arkani-Hamed:2015vfh}
and the references therein),
making it a valuable research topic.

In this work, as a case study, we consider an effective field theory (EFT) based on a non-linear
$\sigma$ model ($\NLSM$), where the Higgs boson arises as a field expansion in
the EFT. The corresponding Higgs couplings to itself and the gauge bosons thus
can be described by their SM couplings modified by some multiplicative
parameters and might very well take non-SM values. As a consequence, 
the vector boson scattering  can be highly enhanced in such classes of models
due to the lack of unitarity cancellations at
high energies. We investigate such an effect with the focus on triple Higgs boson production in 
vector boson fusion process (VBF) at high energy future 
proton-proton colliders. It was shown previously 
that triple Higgs boson production in VBF is especially interesting, since its cross section
increases considerably faster (in comparison to the SM) than for other processes with 
two or three vector bosons in the final states~\cite{Belyaev:2012bm}.
We have found that VBF triple Higgs boson production can only be visible at the 100 TeV $pp$ FCC,
however, the potential of this collider to explore the Higgs coupling to vector bosons 
($hVV$) via this process is impressive:
the process is effectively free of background for the boosted triple Higgs signature
and at high luminosity, the $hVV$ coupling can be measured up to permille precision.

The paper is organised as follows. 
In Section~\ref{sec:unitarity}, we discuss the non-linear $\sigma$ model and unitarity as well as the cross section enhancement 
for multi-boson production in vector boson scattering at high energies.
In Section~\ref{sec:results}, we present results for the signal rates and distributions at the LHC and future $pp$ colliders.
In Section~\ref{sec:background}, we estimate the background for the VBF triple Higgs boson signal and find the potential of the 
100 TeV $pp$ FCC to measure the $hVV$ coupling.
Finally, we draw our conclusions in Section~\ref{sec:conclusion}.

\section{\texorpdfstring{Unitarity and the Non-linear $\sigma$ Model}{Unitarity and the Non-linear Sigma Model}}
\label{sec:unitarity}
In particle physics, one important quantity to describe particle scatterings are their cross sections, and high cross sections mean more likely detections of such scatterings. However, cross sections cannot grow arbitrarily large and are limited by an upper bound, the unitarity bound. For a $2 \to n$ scattering with collision energy $s$, the unitarity bound takes on the form\cite{Dicus:2004rg,Maltoni:2001dc}\\
\begin{equation}
  \sigma (2 \rightarrow n) < \frac{4 \pi}{s}\,.
 \label{eq:unitarity-bound}
\end{equation}
The most general cross section for a $2 \to n$ scattering is proportional to
\begin{equation}
\sigma (2 \to n) \sim \frac{1}{s} \, \mathcal{A}^2(s) \, s^{n-2} \,,
 \label{eq:crossx-uni}
\end{equation}
where $\frac{1}{s}$ corresponds to the flux factor, $\mathcal{A}^2(s)$ is the squared scattering amplitude and $s^{n-2}$ gives the energy dependence of the phase space integral \cite{Dicus:2004rg,Byckling:1973abc}
\begin{equation}
  R_n(s) = \int \prod\limits_{i=1}^{n} \frac{\d^3 p_i}{(2 \pi)^3 \, (2 E_i)^3} \, (2 \pi)^4 \, \delta^4\(\sqrt{s} - \sum\limits_{i=1}^{n} p_i\) = \frac{ (2 \pi)^{4-3n} (\frac{\pi}{2})^{n-1}}{(n-1)! \, (n-2)!} \, s^{n-2}
  \label{eq:dlips}
\end{equation}
for massless particles in four dimensions. Together with Eq. \ref{eq:unitarity-bound}, this restricts the scattering amplitude $\mathcal{A}$ to be proportional to
\begin{equation}
\mathcal{A}(2\to n) \sim s^{1-\frac{n}{2}}
 \label{eq:amp-uni}
\end{equation}
in order for unitarity to be fulfilled.

So far, all considerations have been model-independent, although it turns out that Eq. \ref{eq:amp-uni} is true for the SM, if the theory contains a Higgs boson. {This 
feature of the SM amplitude is special and not generic for 
other models}. Consider the following Lagrangian of a non-linear $\sigma$ model ($\NLSM$)
\begin{equation}
\mathcal{L}_{\NLSM} = \frac{v^2}{4} \, \text{Tr} \[ \del_\mu U \del^\mu U^\dagger \] \,,
\label{eq:NLSM-L}
\end{equation}
where $v=246$ GeV is the usual scale of electroweak symmetry breaking and
\begin{equation}
U = \e^{\frac{\mi \vec{\tau} \cdot \vec{\pi}}{v}}\,.
\label{eq:NLSM-U}
\end{equation}
with $\vec{\pi}$ being the massless Goldstone bosons of the theory. By using the equivalence theorem \cite{Cornwall:1974km,Lee:1977eg,Chanowitz:1985hj,Willenbrock:1987xz,Bagger:1989fc,Veltman:1989ud}, these can be identified with the longitudinal vector bosons in the high energy limit.

One can show by na\"{\i}ve power counting that the scattering amplitudes in the $\NLSM$ grow lienarly in $s$, i.e.
\begin{equation}
  \mathcal{A}_{\NLSM}(2 \rightarrow n) \sim \frac{s}{v^n} \,.
  \label{eq:NLSM-Amp}
\end{equation}
As a consequence, the cross sections grow arbitrarily large and unitarity is violated for any scattering process in the $\NLSM$. In order to restore unitarity, the model must be repaired in the UV region, where unitarity violation occurs\footnote{When the UV region starts varies and depends mainly on how many particles are produced in the final state and how big $s$ is \cite{Belyaev:2012bm}.}. This can be achieved by adding a scalar field, call it the Higgs field, to the model, coupling to the lightest degrees of freedom. This is similar to the case in the SM, where the Higgs is mandatory to cancel unitarity-violating contributions in longitudinal vector boson scattering ($W_LW_L \to W_LW_L$). %However, it is still unknown whether the Higgs found in 2012 at the LHC is the SM Higgs or a different one. For the latter case, the Higgs might have couplings different to the SM Higgs coupling. This is realised in a variety of BSM theories, e.g. in composite Higgs models.

It is convenient to describe the $\NLSM$ together with the Higgs field in terms of an EFT, where we expand operators around the Higgs field. The corresponding Lagrangian takes the following form\cite{Giudice:2007fh}

\begin{align}
  \cal{L}_{\text{eff}}  &= \frac{v^2}{4} \(1 + 2 a \frac{h}{v} + b \frac{h^2}{v^2} + b_3 \frac{h^3}{v^3} + \cdots \) \mbox{Tr} \[ \partial_\mu U \partial^\mu U^\dagger \] \notag \\
  &+ \frac{1}{2} (\partial_\mu h)^2 - \frac{1}{2} m_h^2 h^2  - d_3 \lambda v h^3 - d_4 \frac{\lambda}{4} h^4 + \cdots \,,
  \label{SILH}
\end{align}
where $a,b,b_3,d_3,d_4$ are dimensionless parameters changing the overall coupling strength of a certain term. By setting $a=b=d_3=d_4=1$, $b_3=0$ and redefining the Higgs, the SM is restored and there is no unitarity violation. On the other hand, changing these parameters will lead to large increases in cross sections at high scattering energies along with unitarity violation, since the cancellations mentioned earlier cannot be fully compensated for by the Higgs. The energy scale at which unitarity violation starts to appear therefore is the upper limit for the validity of an EFT. Beyond this scale, the EFT si no longer a good approximation of nature and a new model or modifications to the old one are needed. In either case, this behaviour can be used as an indicator of New Physics.
%{\color{red}XXX(add more about why this is a hint for new physics)} 

In this work, we choose $b=d_3=d_4=1$ and $b_3=0$, but leave $a$ as a free parameter. In other words, we consider the SM with a modified coupling between one Higgs and two gauge bosons.

In the next section, we revisit and update the cross sections for different scattering processes involving the modified couplings, as previously studied in Ref. \cite{Belyaev:2012bm}.

\section{Triple Higgs boson prouction via VBF}
\label{sec:results}
\subsection{\texorpdfstring{Cross sections for multiple vector boson and Higgs production with two jets}{Cross sections for vector boson and Higgs production with two jets}}
In order to estimate which process benefits most of the changed couplings in the EFT, we investigate the process $pp \to jj + X$ with the final states $X$ being either $W^+W^-$, $W^+W^-h$, $hh$ or $hhh$. We compute the regular SM cross sections ($a = 1$) and the cross sections for $a = 0.9$. We further compute these cross sections with applied vector boson fusion (VBF) cuts in order to enhance the actual Higgs signal. The cross sections are computed using \texttt{Madgraph5\_aMC\@NLO 2.2.3} \cite{Alwall:2014hca}. The parton density function (PDF) we use is \texttt{CTEQ6l1} \cite{Pumplin:2002vw}. To avoid soft and collinear jets, we assign a general minimum transverse momentum of the jets to be $p^j_{\rm{T}} \geq 50$ GeV and the minimal distance between two jets is set to $\Delta R(j,j) \geq 0.4$. The proton and jet particle content is set to $(p,j)={g,u,\bar{u},d,\bar{d},s,\bar{s},c,\bar{c}}$. The VBF cuts we choose are listed in Tab.~\ref{tab:VBFcuts}. Finally, the computed cross sections are shown in Tab.~\ref{tab:Madgraph-gudsc}.

\begin{table}[htb]
  \centering
  \begin{tabular}{ccc}
    \toprule
    Parameter & without VBF cuts & with VBF cuts \\
    \midrule
    $E_j$ [GeV]   & 0 & 1500 \\
    $\Delta \eta$ & 0 & 5 \\
    \bottomrule
  \end{tabular}
  \caption{Parameter values with and without VBF cuts.}
  \label{tab:VBFcuts}
\end{table}

%\begin{table}[tbp]
%	\centering
%	  \resizebox{\textwidth}{!}{%
%		\begin{tabular}{cccccccc}
%			\toprule
%			\multirow{2}{*}{Process} & \multirow{2}{*}{VBF cuts} & \multicolumn{2}{c}{13 TeV} & \multicolumn{2}{c}{33 TeV} & \multicolumn{2}{c}{100 TeV} \\\cline{3-8} & & $a=1.0$ & $a=0.9$ & $a=1.0$ & $a=0.9$ & $a=1.0$ & $a=0.9$ \\ \midrule \midrule
%\multirow{2}{*}{$pp \to jjW^+W^-h$} & $\times$ & (-) & (-) & (-) & (-) & (-) & (-)  \\
%								  & \checkmark & - & - & - & - & - & - \\ \midrule          
%\multirow{2}{*}{$pp \to jjhh$} & $\times$ & (-) & (-) & (-) & (-) & (-) & (-) \\
%								  & \checkmark & - & - & - & - & - & - \\ \midrule
%\multirow{2}{*}{$pp \to jjhhh$} & $\times$ & $1.72\times 10^{-4}$ & $2.43 \times 10^{-2}$ & $1.54 \times 10^{-3}$ & $1.18$ & $8.09 \times 10^{-3}$ & $42.35$ \\
%							   & \checkmark & $1.55 \times 10^{-5}$  & $5.66 \times 10^{-3}$ & $5.55 \times 10^{-4}$ & $0.87$ & $3.08 \times 10^{-3}$ & 36.04 \\ \bottomrule 
%
%		\end{tabular}}
%		\caption{Cross sections in fb for different processes at different parameter sets. The proton and jet contents were set to $(p,j)={u,\bar{u},d,\bar{d},s,\bar{s}}$.}
%		\label{tab:CalcHEP-uds}
%\end{table}

\begin{table}[htb]
	\centering
	\resizebox{\textwidth}{!}{%
	\begin{tabular}{cccccccc}
	    \toprule
		\multirow{2}{*}{Process} & \multirow{2}{*}{VBF cuts} & \multicolumn{2}{c}{13 TeV} & \multicolumn{2}{c}{33 TeV} & \multicolumn{2}{c}{100 TeV} \\\cline{3-8}
                                             &            & $a=1.0$              & $a=0.9$              & $a=1.0$              & $a=0.9$              & $a=1.0$              & $a=0.9$             \\ \midrule \midrule
        \multirow{2}{*}{$pp \to jjW^+W^-$}   & $\times$   & 9.88                 & 9.88                 & 60.56                & 60.48                & 352.14               & 352.49               \\
								            & \checkmark & $1.29 \cdot 10^{-2}$ & $1.27 \cdot 10^{-2}$ & 0.48                 & 0.47                 & 5.49                 & 5.47                 \\ \midrule   
        \multirow{2}{*}{$pp \to jjW^+W^-h$}  & $\times$   & $1.71 \cdot 10^{-3}$ & $1.43 \cdot 10^{-3}$ & $1.63 \cdot 10^{-2}$ & $1.53 \cdot 10^{-3}$ & 0.69                 & 0.60                 \\
								            & \checkmark & $1.26 \cdot 10^{-5}$ & $1.35 \cdot 10^{-5}$ & $9.30 \cdot 10^{-4}$ & $1.05 \cdot 10^{-3}$ & 0.15                 & 0.19                 \\ \midrule          
        \multirow{2}{*}{$pp \to jjhh$}       & $\times$   & $5.11 \cdot 10^{-4}$ & $3.64 \cdot 10^{-4}$ & $3.49 \cdot 10^{-3}$ & $2.93 \cdot 10^{-3}$ & $1.70 \cdot 10^{-2}$ & $1.92 \cdot 10^{-2}$ \\
								            & \checkmark & $2.13 \cdot 10^{-5}$ & $1.32 \cdot 10^{-5}$ & $7.65 \cdot 10^{-4}$ & $7.69 \cdot 10^{-4}$ & $5.56 \cdot 10^{-3}$ & $9.20 \cdot 10^{-3}$ \\ \midrule
        \multirow{2}{*}{$pp \to jjhhh$}      & $\times$   & $2.38 \cdot 10^{-7}$ & $2.50 \cdot 10^{-5}$ & $1.97 \cdot 10^{-6}$ & $1.37 \cdot 10^{-3}$ & $1.23 \cdot 10^{-5}$ & $4.60 \cdot 10^{-2}$ \\
							                & \checkmark & $6.14 \cdot 10^{-9}$ & $2.06 \cdot 10^{-6}$ & $4.39 \cdot 10^{-7}$ & $7.48 \cdot 10^{-4}$ & $4.70 \cdot 10^{-6}$ & $4.10 \cdot 10^{-2}$ \\ \bottomrule 
        
		\end{tabular}}
	\caption{Cross sections in pb for different processes with variable $a$, $\sqrt{s}$ and VBF cuts.
	The cross ($\times$) indicates the cross sections before VBF cuts,
	while the tick ($\checkmark$) refers to the cross sections after VBF cuts.}
	\label{tab:Madgraph-gudsc}
\end{table}

The first thing to notice is that all cross sections increase with energy. In the SM case ($a=1$), all cross sections roughly grow by two to three orders of magnitude, if $\sqrt{s}$ is increased from 13 TeV to 100 TeV. This is also true if VBF cuts are applied, albeit the impact of the cuts is very different for different processes. For the first two processes 
with $W^+W^-$ in the final state, VBF cuts will reduce the cross sections by 2 to 3 orders of magnitude, whereas in case of  pure Higgs production channels, the cross sections decreases by a factor of around 30 for 13 TeV and by only a factor around 3 for 100 TeV
collider. The reason for this is that the processes with only Higgs bosons and jets in the final state are mainly produced through VBF, whereas the processes with $W^\pm$ pairs in the final states can be produced through a variety of different channels (e.g. radiation from  jets).

Coming now to the non-SM case ($a=0.9$), triple Higgs production clearly stands out compared to the other processes. Not only it is  least affected by VBF cuts (the cross sections decrease by a factor of 12 at 13 TeV, 1.8 at 33 TeV and 1.1 at 100 TeV), but it also the most significantly enhanced
by the change from $a=1$ to $a=0.9$. At 13 TeV, the cross section after VBF cuts increases by a factor of almost 400, whereas at 100 TeV, the cross sections is almost $10^4$ times larger compared to its SM value. All other processes gain or lose only a negligible part of their SM cross sections. This can be explained by a 
transversal `pollution' of the cross sections, which is highly present in the non-Higgs processes. Here, the transversal contribution to the cross sections is significantly larger compared to the longitudinal part. This also explains why there is no gain in cross section when moving from $a=1$ to $a=0.9$ for $W^+W^-$ production, as there are only very few diagrams actually involve a coupling of two $W^\pm$ to a longitudinal W-bosons.

To summarise the properties of the processes discussed above, it becomes apparent that triple Higgs production offers great potential to explore Higgs properties such as its couplings to other bosons and itself. Due to the huge increase in cross sections (and possible unitarity violations) in the non-SM case, it may also serve as a great tool to explore physics beyond the SM close to the cut-off energy scale of the underlying EFT, as was discussed in chapter~\ref{sec:unitarity}. 
One may not forget, however, that the cross sections for triple Higgs production after all are still only in the range of several fb and small compared to cross sections other processes can achieve. For this reason, it is also important to investigate the possible backgrounds for triple Higgs production, estimate the signal-to-background ratio and a signal significance for a given luminosity. This analysis is performed in chapter~\ref{sec:background}.

For the following part, we focus solely on triple Higgs production with applied VBF cuts, and study the impact of the anomalous Higgs coupling $a$ for  different collision energies and unitarity bounds.

\subsection{Vector boson scattering level and Unitarity}
\begin{figure}[h]
  \begin{center}
    \includegraphics[scale=.4]{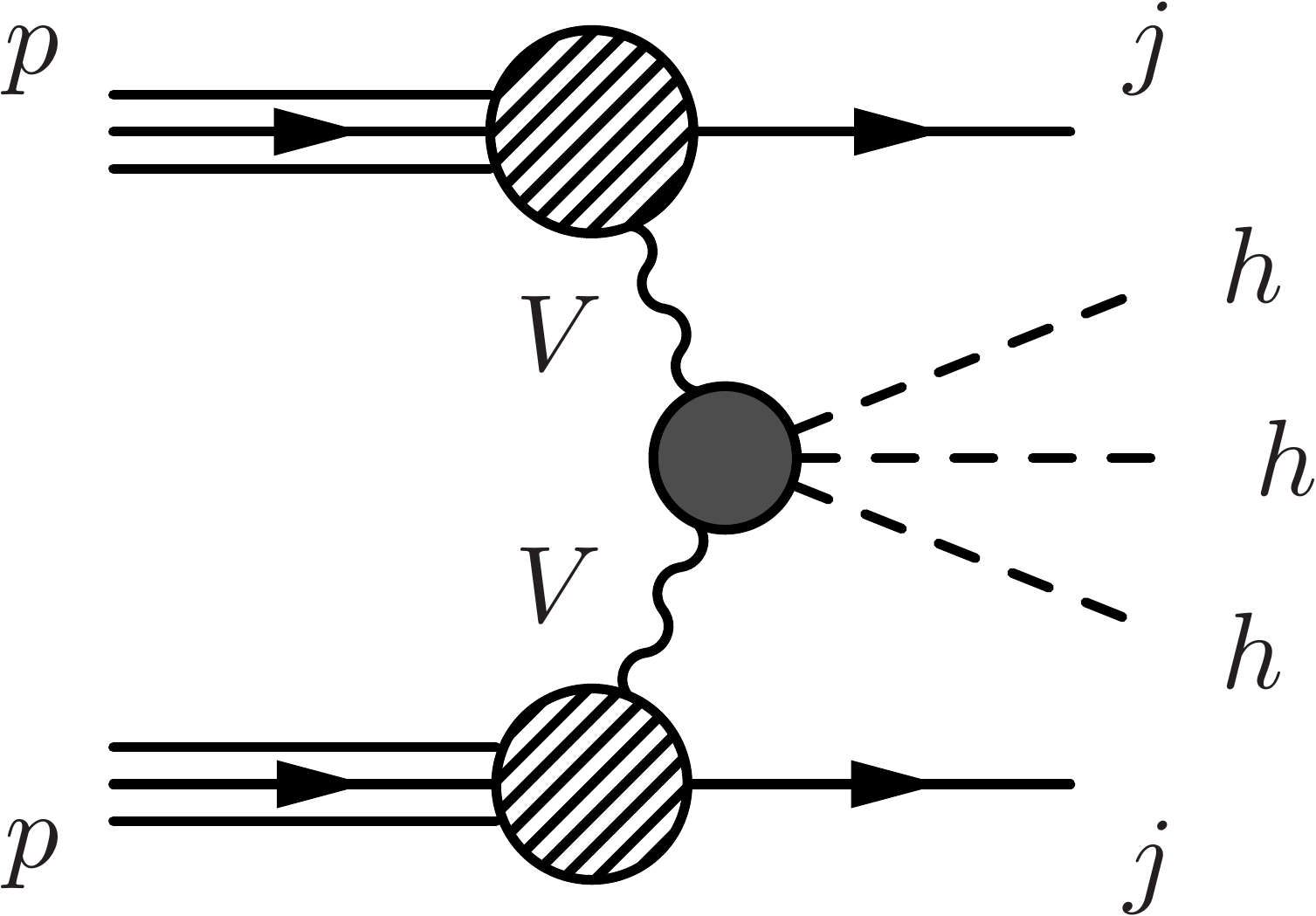}
    \caption{Schematic diagram for triple Higgs production in VBF. The grey blob in the centre represents many Feynamn diagrams and topologies for two vector-bosons $V=Z,W^\pm$ fusion
    into  three Higgs bosons $h$.}
    \label{fig:feynman-pp_to_jjhhh}
  \end{center}
\end{figure}

In Fig.~\ref{fig:feynman-pp_to_jjhhh} we present a schematic diagram for triple Higgs production,
 which represents the process under study and the around a hundred actual Feynman diagrams behind it.
Before calculating the cross sections for the full hadronic process, however, it is worth investigating only the VBF part of this process, i.e. $VV \to hhh$ with $V=Z,W^\pm$.
In this case, the invariant mass of the three Higgs bosons $M_{hhh}$ is equal to the $VV$ center-of-mass (CM) energy $\sqrt{\hat{s}_{\text{VBF}}}$, so
\begin{equation}
  M_{hhh} = \sqrt{\hat{s}_{\text{VBF}}} \,.
  \label{eq:mhhh=sqrts}
\end{equation}
This relation is very useful in two ways. First, it can be used to calculate the unitarity bound at the VBF stage with high precision. This is achieved by plugging in the cross sections for  $VV \to hhh$, $\sigma_{VV \to  VVhhh} \equiv \hat\sigma(hhh)$, in Eq.~\ref{eq:unitarity-bound} and solving for $\sqrt{\hat{s}_{\text{VBF}}}$, which now marks the CM energy, where unitarity is violated. Second, it acts as a link between the  level of $VV$ scattering 
and $qq$ scattering. So if parts of this distribution exceed the unitarity bound found in $\sqrt{\hat{s}_{\text{VBF}}}$, this clearly indicates the presence of New Physics,
in particular some resonances which should unitarise the scattering amplitude.

\begin{figure}[htb]
  \begin{center}
    %\fontsize{17.28}{12}\selectfont 
    \includegraphics[width=0.8\textwidth]{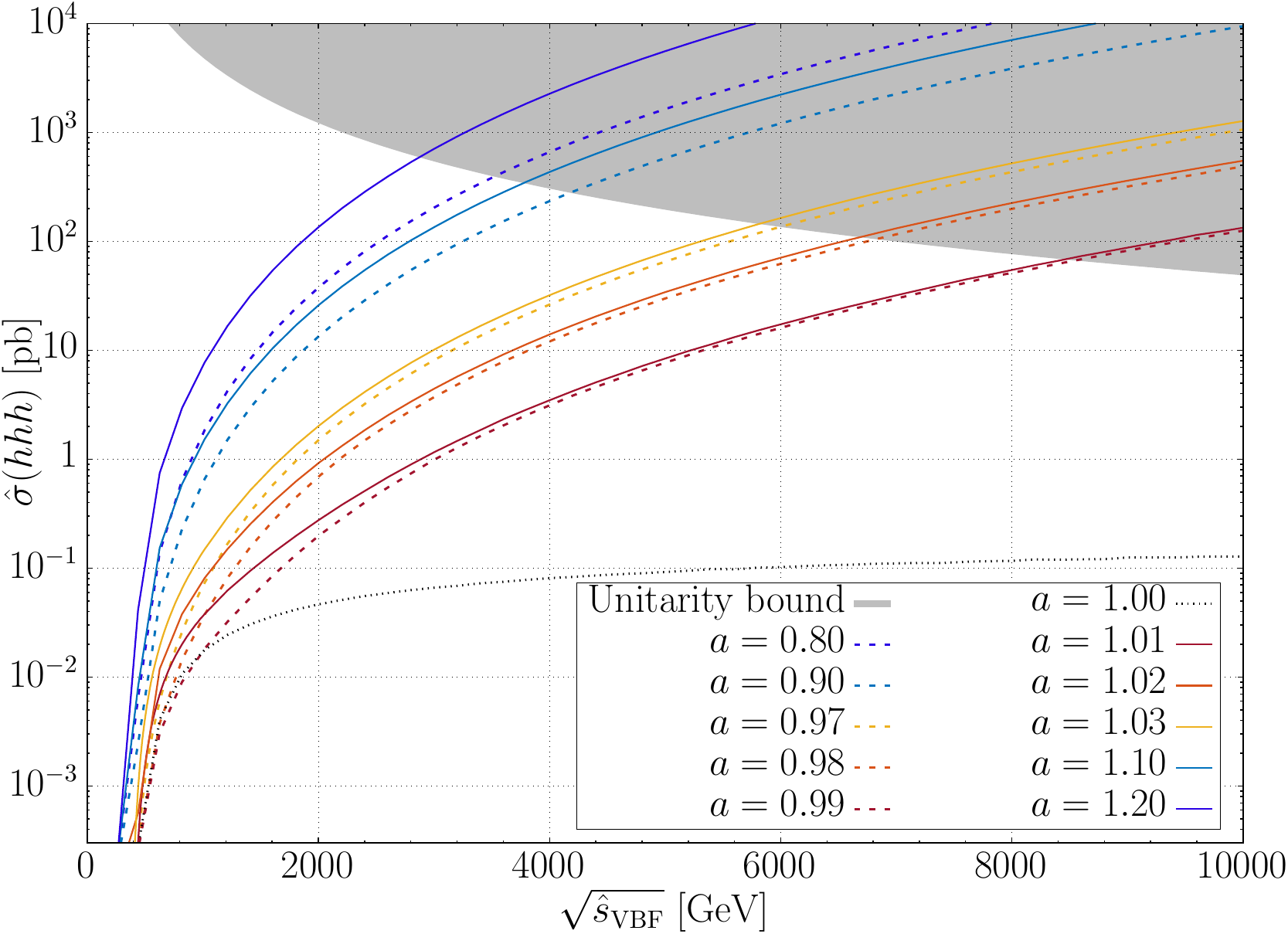}
    \caption{Cross sections $\hat\sigma(hhh)$ in pb for vector boson scattering into three Higgs, $VV \to hhh$, $V = Z,W^\pm$, for different values of $a$. The grey area marks the region where unitarity is violated.}
    \label{fig:vvtohhh}
  \end{center}
\end{figure}

In order to address the first point, we computed the cross sections for $VV \to hhh$ and its dependence on $a$ using \texttt{CalcHEP 3.6.23} \cite{Belyaev:2012qa}. Fig.~\ref{fig:vvtohhh} shows a series of these cross sections for different values of $a$ together with the unitarity bound (Eq.~\ref{eq:unitarity-bound}.)
The coloured curves show the cross sections as functions of $\sqrt{\hat{s}_{\text{VBF}}}$, where dashed lines refer to $a < 1$ and the solid curves show the cross sections with $a > 1$. The dotted black line at the bottom shows the SM cross section for comparison and the grey area in the top right corner marks the region where unitarity is violated. 
One can observe a huge increase of the  cross section for any value of $a \neq 1$ compared to the SM, as discussed in chapter~\ref{sec:unitarity}.
For the shown range of $a \neq 1$, unitarity is violated roughly between $\sqrt{\hat{s}_{\text{VBF}}} = 2.4$ TeV and 8.4 TeV
for $|a-1|$ range between 0.2 and 0.01 respectively.
Eventually for $a=1$ unitarity is not  violated, since there is no unitarity violation in the SM.

\subsection{Differential distributions }
\begin{figure}[htb]
  \begin{center}
    \includegraphics[width=\textwidth]{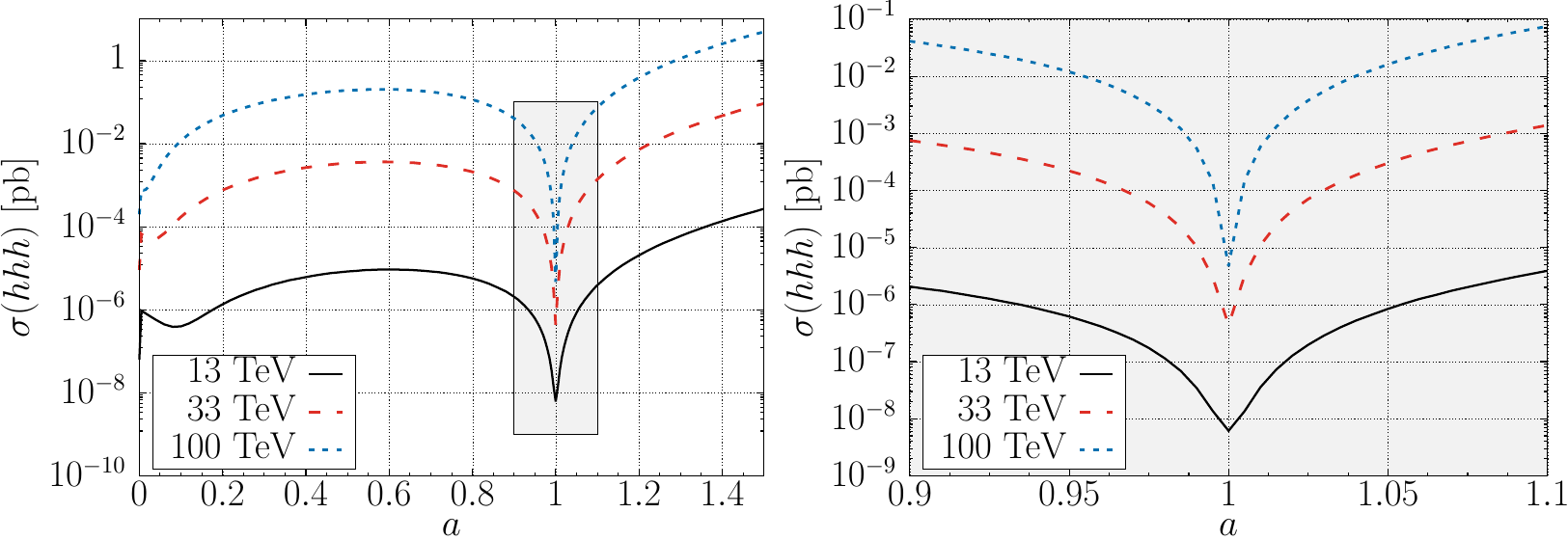}
    \caption{Cross sections $\sigma(hhh)$ in pb for $p p \to j j h h h$ with VBF cuts for $\sqrt{s} = $ 13, 33, 100 TeV in dependance of $a$. The right plot shows a zoomed version of the grey highlighted segment with $a \in [0.9,1.1]$ in the left plot.}
    \label{fig:both-sigma-a}
  \end{center}
\end{figure}

In the last chapter, we have seen that triple Higgs production is greatly enhanced 
when the $hVV$ coupling deviates from the SM one even at the percent level. 
In this section we take a closer look at the $pp \to jjhhh$, $\sigma_{pp \to jjhhh} \equiv \sigma(hhh)$ cross section as a function of anomalous $hVV$ coupling $a$, unitarity violation and differential distribution
of the $hhh$ invariant mass. For this purpose  using \texttt{Madgraph5\_aMC\@NLO 2.2.3} we computed the total  cross section for $pp \to jjhhh$, $\sigma_{pp \to jjhhh} \equiv \sigma(hhh)$ with vector-boson fusion (VBF) cuts applied
as a function of $a$  parameter. The results are shown in Fig.~\ref{fig:both-sigma-a}.
For $a=1$, the SM coupling is restored and therefore also the SM cross section. However, even for 
a small deviation of $a$ from one
e.g. for $a=0.98$, the cross sections increase by more than one order of magnitude for $\sqrt{s} = 13$ TeV,
by more than two orders of magnitude for $\sqrt{s} = 33$ TeV and
by about three orders of magnitude for $\sqrt{s} = 100$ TeV. If $a$ deviates roughly 10 \% from 1, the increase starts to slow down. For even smaller values of  $a$ the cross section reaches extremum at $a \approx 0.6$ and by even further reducing $a$, the cross sections starts to decrease again. On the other side, increasing $a$ beyond 1 will lead only to ever growing cross sections, as the multiplicative nature of $a$ in the coupling starts to dominate the cross sections slope. This behaviour has been studied and explained in~\cite{Belyaev:2012bm}
at the level of $WW$ scattering and is well reproduced here at the level of pp collisions.

\begin{figure}[htb]
  \begin{center}
    \includegraphics[width=\textwidth]{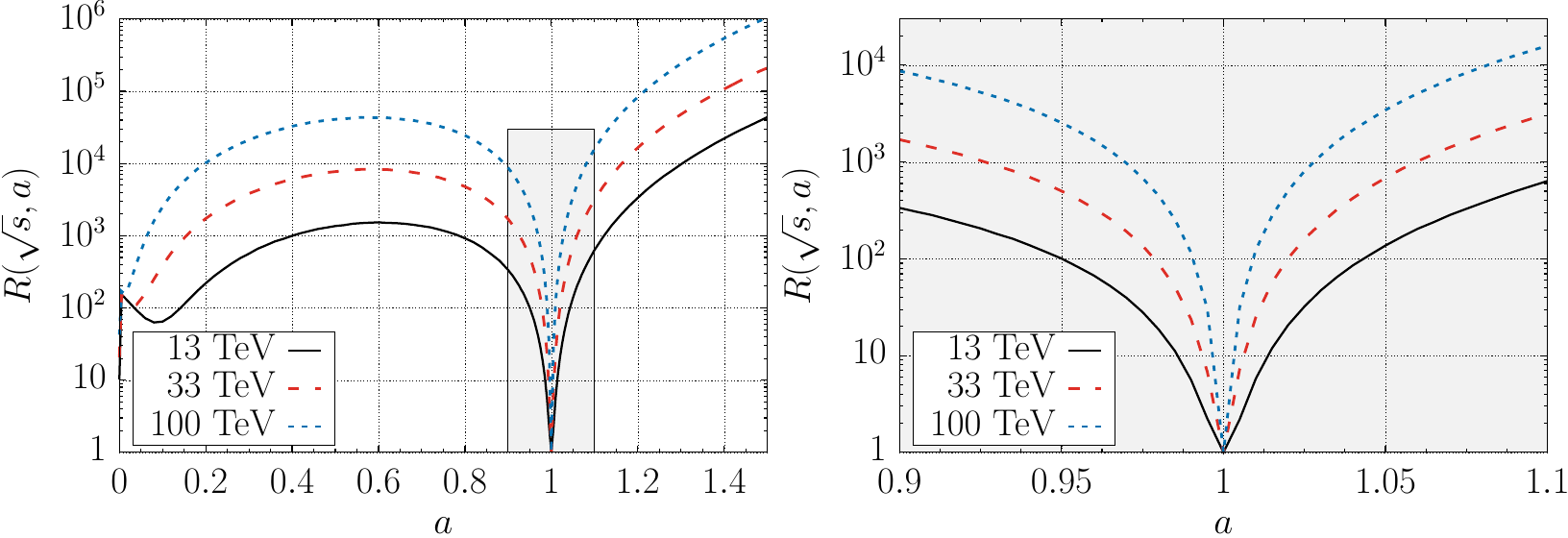}
    \caption{Ratio $R_{\sqrt{s}}(a) = \frac{\sigma^{pp \to jjhhh}(a)}{\sigma^{pp \to jjhhh}(a=1)}$ for $pp \to jjhhh$ with the data of Fig.~\ref{fig:both-sigma-a}. $R_{\sqrt{s}}(a) = 1$ corresponds to the unmodified SM cross sections. The right plot shows a zoomed version of the grey highlighted segment with $a \in [0.9,1.1]$ in the left plot.}
    \label{fig:both-ratio}
  \end{center}
\end{figure}

In order to compare the cross sections for different $\sqrt{s}$ and to see the actual gain in cross section compared to the SM, it is useful to normalise the data of Fig.~\ref{fig:both-sigma-a} with respect to the SM cross section $\sigma(a=1)$. The resulting cross sections are shown in Fig.~\ref{fig:both-ratio}. 
Again, the cross section increases fastest in the area $a \in [0.9,1.1]$.
This huge enhancement in cross section however comes with the price of (partially) losing unitarity, since there is no exact Higgs cancellation in the VBF channel any longer. As this loss of unitarity indicates where new physics must appear, it is important to know at which energies  unitarity is violated and how distinct the violation is.

\begin{figure}[htb]
  \begin{center}
    %\fontsize{17.28}{12}\selectfont 
    \includegraphics[width=0.7\textwidth]{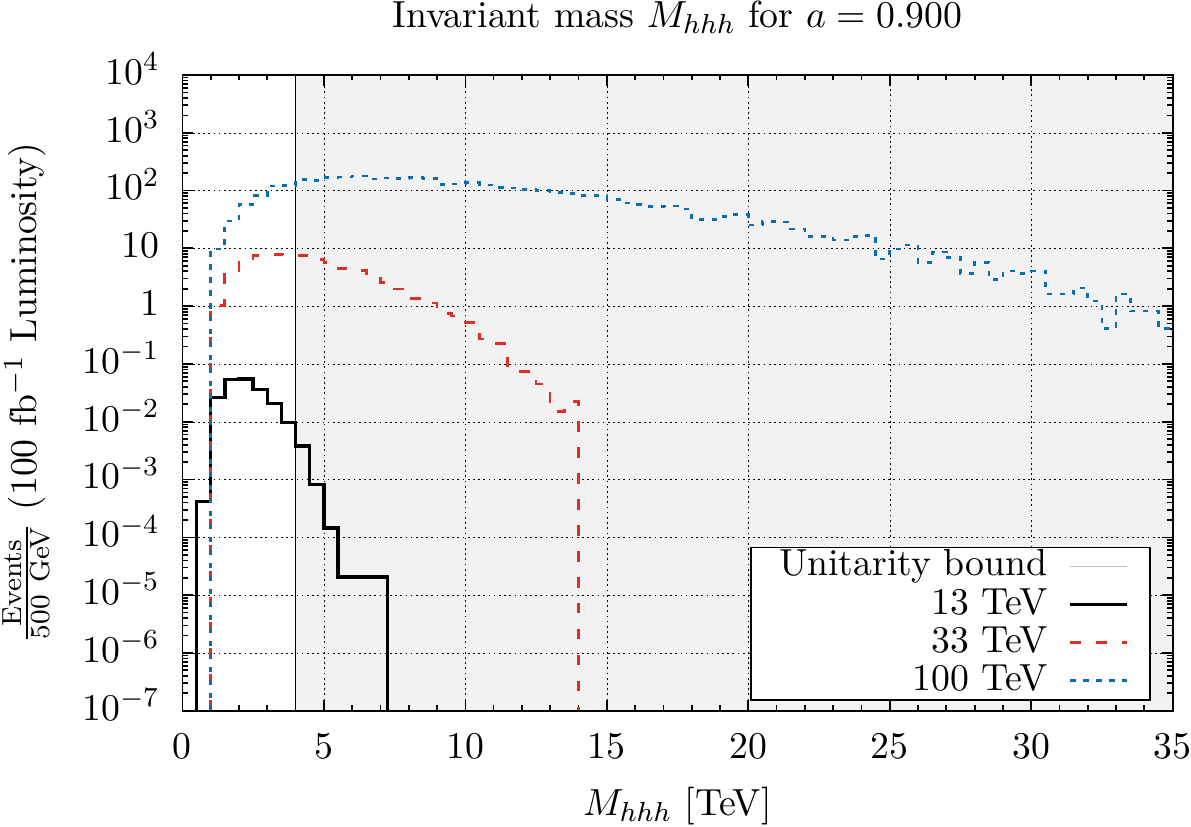}
    \caption{Invariant mass $M_{hhh}$ in the process $p p \to j j h h h$ at $a=0.9$ for $\sqrt{s} = 13, 33, 100$ TeV. The shaded area marks the region where unitarity is violated.}
    \label{fig:M_hhh_0.9}
  \end{center}
\end{figure}

\begin{figure}[htb]
  \begin{center}
    %\fontsize{17.28}{12}\selectfont 
    \includegraphics[width=0.7\textwidth]{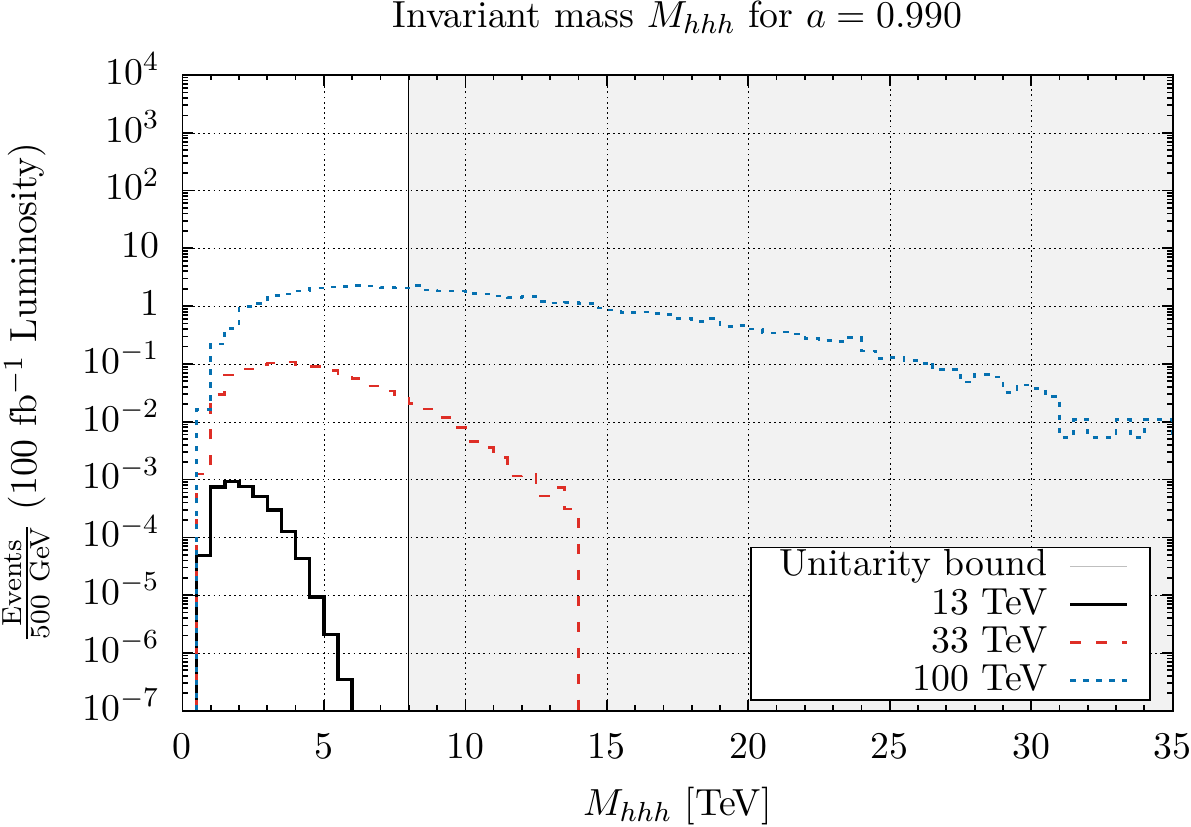}
    \caption{Invariant mass $M_{hhh}$ in the process $p p \to j j h h h$ at $a=0.99$ for $\sqrt{s} = 13, 33, 100$ TeV. The shaded area marks the region where unitarity is violated.}
    \label{fig:M_hhh_0.99}
  \end{center}
\end{figure}

Since  the unitarity violating energy scales of $\sqrt{\hat{s}_{\text{VBF}}}$ are known, we can apply this knowledge for $pp \to jjhhh$  at the level of pp collisions. To do so, we computed the full invariant mass $M_{hhh}$ for $pp \to jjhhh$ using \texttt{Madgraph5\_aMC\@NLO 2.2.3} to generate the events, and \texttt{ROOT 5.34.25} \cite{Brun:1996abc} to obtain the invariant mass distributions. Fig.~\ref{fig:M_hhh_0.9} and~\ref{fig:M_hhh_0.99} show the full invariant mass $M_{hhh}$ in TeV for two representative values of $a$ used in Fig~\ref{fig:vvtohhh}. The grey area again marks the region where unitarity is violated.

The invariant mass distributions all appear very similar with respect to $a$ and have their peaks around 1.8 TeV for $\sqrt{s} = 13$ TeV, 3.5 TeV for $\sqrt{s} = 33$ TeV and 7 TeV for $\sqrt{s} = 100$ TeV. However, with increasing $\sqrt{s}$, the distributions smear out and the tail at high $M_{hhh}$ gets longer
and flatter reflecting non-unitary behaviour of the amplitude with high $M_{hhh}$. Also one can  notice how the unitarity bound shifts to higher energies if $a$ approaches 1, as seen in Fig.~\ref{fig:vvtohhh}.
At $a=0.9$, unitarity violation roughly starts at $M_{hhh} = 4$ TeV
while for $a=0.99$, the unitarity bound is at around 8 TeV. 
To indicate the proportion of scattering events that violate unitarity for each value of $a$, we define a parameter, $\mathcal{U}$, as:
\begin{equation}
  \mathcal{U} = \frac{\# \text{ (events violating unitarity)}}{\# \text{ (all events)}} \,.
 \label{eq:unitarity-violation}
\end{equation}
We also define $\varepsilon_{\mathcal{U}} = 1 - \mathcal{U}$, which gives the proportion of scattering events where unitarity is not violated.
Tab.~\ref{tab:Mhhh} shows the fraction of events (or fraction of differential cross section)  not violating unitarity with respect to $a$ and $\sqrt{s}$, where we assumed the total integrated luminosity to be $\mathcal{L}_{\text{int}} = 100$ fb$^{-1}$ for all energies. This allows easy comparisons between the energies and other references.

\begin{sidewaystable}
	\centering
      \resizebox{\textwidth}{!}{%
		\begin{tabular}{ccccccccccccccc}
		\toprule
        \multicolumn{5}{c}{13 TeV} & \multicolumn{5}{c}{33 TeV} & \multicolumn{5}{c}{100 TeV} \\\midrule
        \multicolumn{1}{c|}{$a$}  & $\varepsilon_{\mathcal{U}}$ [\%]  & $\sigma$ [pb]          & $\mathcal{L}_{\text{int}} \cdot \sigma$  & \multicolumn{1}{c||}{$\mathcal{L}_{\text{int}} \cdot \sigma \cdot \varepsilon_{\mathcal{U}}$}  & \multicolumn{1}{c|}{$a$}   & $\varepsilon_{\mathcal{U}}$ [\%] & $\sigma$ [pb]                  & $\mathcal{L}_{\text{int}} \cdot \sigma$ & \multicolumn{1}{c||}{$\mathcal{L}_{\text{int}} \cdot \sigma \cdot \varepsilon_{\mathcal{U}}$} & \multicolumn{1}{c|}{$a$}  & $\varepsilon_{\mathcal{U}}$ [\%] & $\sigma$ [pb] & $\mathcal{L}_{\text{int}} \cdot \sigma$ & $\mathcal{L}_{\text{int}} \cdot \sigma \cdot \varepsilon_{\mathcal{U}}$ \\\midrule
        %\multicolumn{1}{c|}{0.70}    &    96.19     &    $8.40 \cdot 10^{-6}$    &    0.84    &    \multicolumn{1}{c||}{0.81}    &    \multicolumn{1}{c|}{0.70}    &    39.81     &    $3.20 \cdot 10^{-3}$    &    320.00     &    \multicolumn{1}{c||}{127.39}    &    \multicolumn{1}{c|}{0.70}    &    8.48      &    0.18                    &    18000.00     &    1526.40  \\
        \multicolumn{1}{c|}{0.80}    &    97.81     &    $5.70 \cdot 10^{-6}$    &    0.57    &    \multicolumn{1}{c||}{0.56}    &    \multicolumn{1}{c|}{0.80}    &    44.76     &    $2.10 \cdot 10^{-3}$    &    210.00     &    \multicolumn{1}{c||}{94.00}     &    \multicolumn{1}{c|}{0.80}    &    10.01     &    0.12                    &    12000.00     &    1201.20  \\
        \multicolumn{1}{c|}{0.90}    &    99.72     &    $2.10 \cdot 10^{-6}$    &    0.21    &    \multicolumn{1}{c||}{0.21}    &    \multicolumn{1}{c|}{0.90}    &    58.21     &    $7.50 \cdot 10^{-4}$    &    75.00      &    \multicolumn{1}{c||}{43.66}     &    \multicolumn{1}{c|}{0.90}    &    15.38     &    $4.10 \cdot 10^{-2}$    &    4100.00      &    630.58   \\
        \multicolumn{1}{c|}{0.92}    &    99.30     &    $1.40 \cdot 10^{-6}$    &    0.14    &    \multicolumn{1}{c||}{0.14}    &    \multicolumn{1}{c|}{0.92}    &    62.36     &    $5.10 \cdot 10^{-4}$    &    51.00      &    \multicolumn{1}{c||}{31.80}     &    \multicolumn{1}{c|}{0.92}    &    17.48     &    $2.80 \cdot 10^{-2}$    &    2800.00      &    489.44   \\
        \multicolumn{1}{c|}{0.94}    &    99.99     &    $8.50 \cdot 10^{-7}$    &    0.09    &    \multicolumn{1}{c||}{0.08}    &    \multicolumn{1}{c|}{0.94}    &    68.75     &    $3.10 \cdot 10^{-4}$    &    31.00      &    \multicolumn{1}{c||}{21.31}     &    \multicolumn{1}{c|}{0.94}    &    19.84     &    $1.70 \cdot 10^{-2}$    &    1700.00      &    337.28   \\
        \multicolumn{1}{c|}{0.96}    &    100       &    $4.10 \cdot 10^{-7}$    &    0.04    &    \multicolumn{1}{c||}{0.04}    &    \multicolumn{1}{c|}{0.96}    &    76.29     &    $1.50 \cdot 10^{-4}$    &    15.00      &    \multicolumn{1}{c||}{11.44}     &    \multicolumn{1}{c|}{0.96}    &    24.76     &    $7.90 \cdot 10^{-3}$    &    790.00       &    195.60   \\
        \multicolumn{1}{c|}{0.97}    &    100       &    $2.40 \cdot 10^{-7}$    &    0.02    &    \multicolumn{1}{c||}{0.02}    &    \multicolumn{1}{c|}{0.97}    &    82.61     &    $8.40 \cdot 10^{-5}$    &    8.40       &    \multicolumn{1}{c||}{6.94}      &    \multicolumn{1}{c|}{0.97}    &    29.97     &    $4.60 \cdot 10^{-3}$    &    460.00       &    137.86   \\
        \multicolumn{1}{c|}{0.98}    &    100       &    $1.10 \cdot 10^{-7}$    &    0.01    &    \multicolumn{1}{c||}{0.01}    &    \multicolumn{1}{c|}{0.98}    &    89.69     &    $3.90 \cdot 10^{-5}$    &    3.90       &    \multicolumn{1}{c||}{3.50}      &    \multicolumn{1}{c|}{0.98}    &    35.53     &    $2.10 \cdot 10^{-3}$    &    210.00       &    74.61    \\
        \multicolumn{1}{c|}{0.99}    &    100       &    $3.40 \cdot 10^{-8}$    &    0.00    &    \multicolumn{1}{c||}{0.00}    &    \multicolumn{1}{c|}{0.99}    &    96.91     &    $1.00 \cdot 10^{-5}$    &    1.00       &    \multicolumn{1}{c||}{0.97}      &    \multicolumn{1}{c|}{0.99}    &    50.34     &    $5.40 \cdot 10^{-4}$    &    54.00        &    27.18    \\
        \multicolumn{1}{c|}{1.01}    &    100       &    $3.60 \cdot 10^{-8}$    &    0.00    &    \multicolumn{1}{c||}{0.00}    &    \multicolumn{1}{c|}{1.01}    &    96.72     &    $1.10 \cdot 10^{-5}$    &    1.10       &    \multicolumn{1}{c||}{1.06}      &    \multicolumn{1}{c|}{1.01}    &    48.88     &    $5.90 \cdot 10^{-4}$    &    59.00        &    28.84    \\
        \multicolumn{1}{c|}{1.02}    &    100       &    $1.30 \cdot 10^{-7}$    &    0.01    &    \multicolumn{1}{c||}{0.01}    &    \multicolumn{1}{c|}{1.02}    &    88.64     &    $4.50 \cdot 10^{-5}$    &    4.50       &    \multicolumn{1}{c||}{3.99}      &    \multicolumn{1}{c|}{1.02}    &    35.10     &    $2.40 \cdot 10^{-3}$    &    240.00       &    84.24    \\
        \multicolumn{1}{c|}{1.03}    &    100       &    $2.90 \cdot 10^{-7}$    &    0.03    &    \multicolumn{1}{c||}{0.03}    &    \multicolumn{1}{c|}{1.03}    &    81.76     &    $1.00 \cdot 10^{-4}$    &    10.00      &    \multicolumn{1}{c||}{8.18}      &    \multicolumn{1}{c|}{1.03}    &    27.59     &    $5.50 \cdot 10^{-3}$    &    550.00       &    151.75   \\
        \multicolumn{1}{c|}{1.04}    &    99.98     &    $5.30 \cdot 10^{-7}$    &    0.05    &    \multicolumn{1}{c||}{0.05}    &    \multicolumn{1}{c|}{1.04}    &    74.37     &    $1.90 \cdot 10^{-4}$    &    19.00      &    \multicolumn{1}{c||}{14.13}     &    \multicolumn{1}{c|}{1.04}    &    23.16     &    $1.00 \cdot 10^{-2}$    &    1000.00      &    231.60   \\
        \multicolumn{1}{c|}{1.06}    &    99.94     &    $1.30 \cdot 10^{-6}$    &    0.13    &    \multicolumn{1}{c||}{0.13}    &    \multicolumn{1}{c|}{1.06}    &    65.06     &    $4.50 \cdot 10^{-4}$    &    45.00      &    \multicolumn{1}{c||}{29.28}     &    \multicolumn{1}{c|}{1.06}    &    18.07     &    $2.40 \cdot 10^{-2}$    &    2400.00      &    433.68   \\
        \multicolumn{1}{c|}{1.08}    &    99.56     &    $2.30 \cdot 10^{-6}$    &    0.23    &    \multicolumn{1}{c||}{0.23}    &    \multicolumn{1}{c|}{1.08}    &    56.84     &    $8.40 \cdot 10^{-4}$    &    84.00      &    \multicolumn{1}{c||}{47.75}     &    \multicolumn{1}{c|}{1.08}    &    14.94     &    $4.50 \cdot 10^{-2}$    &    4500.00      &    672.30   \\
        \multicolumn{1}{c|}{1.10}    &    99.01     &    $3.90 \cdot 10^{-6}$    &    0.39    &    \multicolumn{1}{c||}{0.39}    &    \multicolumn{1}{c|}{1.10}    &    50.96     &    $1.40 \cdot 10^{-3}$    &    140.00     &    \multicolumn{1}{c||}{71.34}     &    \multicolumn{1}{c|}{1.10}    &    12.12     &    $7.50 \cdot 10^{-2}$    &    7500.00      &    909.00   \\
        \multicolumn{1}{c|}{1.20}    &    91.81     &    $2.00 \cdot 10^{-5}$    &    2.00    &    \multicolumn{1}{c||}{1.84}    &    \multicolumn{1}{c|}{1.20}    &    32.60     &    $7.20 \cdot 10^{-3}$    &    720.00     &    \multicolumn{1}{c||}{234.72}    &    \multicolumn{1}{c|}{1.20}    &    7.04      &    0.39                    &    39000.00     &    2745.60  \\
        %\multicolumn{1}{c|}{1.30}    &    82.66     &    $6.00 \cdot 10^{-5}$    &    6.00    &    \multicolumn{1}{c||}{4.96}    &    \multicolumn{1}{c|}{1.30}    &    22.13     &    $2.10 \cdot 10^{-2}$    &    2100.00    &    \multicolumn{1}{c||}{464.73}    &    \multicolumn{1}{c|}{1.30}    &    4.08      &    1.10                    &    110000.00    &    4488.00  \\
        \bottomrule
		\end{tabular}}
		\caption{Proportion of scattering events where unitarity is not violated $\varepsilon_{\mathcal{U}}$ in \% for $M_{hhh}$ with respect to $a$ and $\sqrt{s}$. Also shown are the cross sections $\sigma$ in pb, the total number of events $\mathcal{L}_{\text{int}} \cdot \sigma$ and the proportion of events not violating unitarity $\mathcal{L}_{\text{int}} \cdot \sigma \cdot \varepsilon_{\mathcal{U}}$, where $\mathcal{L}_{\text{int}}$ is the total integrated luminosity. We assume $\mathcal{L}_{\text{int}} = 100$ fb$^{-1}$ for all energies to allow easy comparisons between the energies.}
		\label{tab:Mhhh}
\end{sidewaystable}

For $\sqrt{s} = 13$ TeV, unitarity is violated by less than 1 \% of scattering events if $a$ is changed by less than $\pm$ 10 \%. The total number of events, however, is vanishingly small for the whole range of $a$ and the number of events violating unitarity is even smaller, making it nearly impossible to detect such a signal.
For $\sqrt{s} = 33$ TeV, $\mathcal{U}$ is still below 4 \% if $a$ differs only by 1 \% from the SM value. Changing $a$ further leads to $\mathcal{U}$ reaching  40\% when $a$ differs by 10\% from the SM. The number of events for $a=1.01$  is about 1.
For $\sqrt{s} = 100$ TeV, $\mathcal{U}$ is always greater than about 50\% for 1\% deviation from SM
and by 85\% for 10 \% from the SM. This clearly indicates that the chosen NL$\sigma$M cannot be valid at or beyond $\sqrt{s} = 100$ TeV, meaning that new physics should become visible at this energy. The total number of events is comparatively high, ranging from around 50 events for $a=1.01$ to more than $4100$ events for $a = 0.9$. 
One can see  in case of 100 TeV collider even for 1\% deviation of $hVV$ coupling from  the SM one can get a non-negligible number of signal events, however, to judge if the signal can be observed or not
we need to estimate the respective  background. This is the subject of the next section.

\section{Estimation of background and collider sensitivity to $hVV$ coupling}
\label{sec:background}
Triple Higgs production via VBF gives {rise to a} spectacular signature at the FCC:
the invariant mass of {the three} Higgs bosons is above several TeV, even for the case when the $hVV$ coupling 
differs from the SM by only 1 \%. This makes the Higgs bosons quite boosted and even for $M_{hhh} \simeq 1$~TeV,
which is the lower edge of the $M_{hhh}$ distribution, as one can see from Fig.~\ref{fig:M_hhh_0.99},
the cone size around the Higgs boson decay products (e.g. two $b$-jets) will be of the order of 
$\frac{M_h}{2}/\frac{M_{hhh}}{3} = \frac{125{\rm GeV}}{2}/\frac{1000{\rm GeV}}{3} \simeq 0.2$.
Therefore, the signature  will be two forward-backward jets with a large rapidity gap and 
three energetic Higgs-jets with a typical radius below 0.2. 
In this study, we consider the $h \to b\bar{b}$ decay channel for all three Higgs bosons.
In Ref.~\cite{Gouzevitch:2013qca}, the authors have found that the efficiency 
for the identification of a pair of boosted Higgs bosons from KK-Graviton decays
(including $b$-tagging efficiencies) is about $\varepsilon_{hh} \simeq 15\%$ for Higgs bosons with large enough momentum.
These important results are very relevant to our study, where we estimate signal and background
rates using this efficiency.
Using $\varepsilon_{hh}$ one can estimate the efficiency for triple Higgs-jet tagging as
$\varepsilon_{hhh} = \left(\sqrt{\varepsilon_{hh}}\right)^3 \simeq 0.058$.
Taking into account that BR$(h \to b\bar{b})\simeq 58 \%$,  the rate 
for the tagged triple Higgs-jet signature coming from the $pp \to jjhhh$ VBF process followed by $h \to b\bar{b}$ decays
is given by
\begin{equation}
	\sigma_{\text{sig}}(hhh) = \sigma(pp \to jjhhh) \times \varepsilon_{hhh} \times \text{BR}(h \to b\bar{b})^3 \simeq \sigma(pp \to jjhhh) \times 0.0113
	\label{eq:sig}
\end{equation}

We assume that the main background (BG) is coming from the QCD process $pp \to jjb\bar{b}b\bar{b}b\bar{b}$ (6$b$ BG).
Before evaluating this process (which is actually not currently possible by means of known matrix-element generators), we have decided to evaluate the $pp \to b\bar{b}b\bar{b}b\bar{b}$ process to understand the level of the 6$b$-jet background first without the requirement of the two forward-backward jets with large rapidity gap.

To evaluate the background to the triple Higgs-jet signature coming from the 6$b$-jet process, we use a mass window cut
\begin{equation}
	|M^i_{bb}-M_{h}| =  \Delta^i_{M_h} \leq \Delta^{\text{cut}}_{M_h} = 15~{\rm GeV} 
	\label{eq:mhcut}
\end{equation}
for $M^i_{bb} \ (i=1,2,3)$, which represents the three `best' $bb$ or $b\bar{b}$ combinations
with the lowest  $\Delta^i_{M_h}$ values. This choice allows to avoid combinatorial BG.
The choice of $\Delta^{\text{cut}}_{M_h}$ (which can be further optimised)
is made  to be consistent with the jet energy resolution, which is below 10 \% at the ATLAS and CMS detectors at the LHC and which is expected to be of the same order at 100 TeV $pp$ FCC's (FCC@100TeV).
We also apply
\begin{equation}
	p_T^b> 50~{\rm GeV}\,, \quad |\eta_b|<2 \qquad \rm{and} \qquad M_{6b}> 1~{\rm TeV} \,,
	\label{eq:etacut}
\end{equation}
where the first two cuts ensure that the $b$-jets are in the acceptance region and 
the last one is used to effectively suppress the BG, which drops steeply with $M_{6b}$, as illustrated below in Fig.~\ref{fig:bg-dist}.
At the same time, $M_{6b}$ for the signal grows with the increase of $M_{6b}$
for $\varepsilon_a = a-1$ in the $10^{-3}$--$10^{-1}$ range and is not visibly affected by this cut.
Besides the above cuts, we also would like to make use of the fact that the Higgs bosons are quite boosted
and therefore apply an {\it upper} cut on the $\Delta R_{bb}$ separation of the $b$-quarks:
\begin{equation}
	\Delta R_{bb}=\sqrt{\Delta \phi_{bb} + \Delta \eta_{bb}} \leq \Delta R^{\text{cut}}_{bb} = 0.5 \,,
	\label{eq:drcut}
\end{equation}
which will not affect the signal but will further suppress the BG as we illustrate below {in Fig.~\ref{fig:bg-dist}}.

There are certain technical problems in the evaluation of 6$b$ BG: the application 
of the $\Delta_{M_h}^{\text{cut}}$ and $M_{6b}$ cuts at \texttt{MadGraph} level in form of user defined cuts lead to zero cross section due to too little phase space left, so \texttt{MadGraph} was failing to evaluate it.
On the other hand, the 6$b$ BG was too heavy for the squared matrix element method 
of \texttt{CalcHEP} to perform the symbolic calculations. However, we still managed to estimate the 6$b$ BG
using the following procedure:
a) we evaluated the process $pp \to b\bar{b}b\bar{b}$ (4$b$ BG) at parton level using \texttt{CalcHEP} with the cuts given by equations~\eqref{eq:mhcut}--\eqref{eq:drcut} and simulated the respective events;
b) we have used these events as a user process for \texttt{PYTHIA 8.2.30}~\cite{Sjostrand:2007gs}
Monte-Carlo generator to find the probability of producing an additional $b\bar{b}$-pair from 
initial (ISR) and final state radiation (FSR) and have applied the kinematical cuts of equations~\eqref{eq:mhcut} -- \eqref{eq:drcut} 
on this pair at \texttt{PYTHIA} level;
c) we validated this procedure for lower $b$-quark multiplicities
(by simulating the 4$b$ BG from a 2$b$ BG starting point) and have found that for ISR/FSR and the QCD scale in \texttt{PYTHIA}
chosen to be equal to $\hat{s}$, this estimation works with an accuracy of about 20--30 \%. One should note that this is sufficient to estimate the 6b BG within an order of magnitude 
as we discuss below.
To illustrate the importance of $M_{4b}$ and $\Delta R_{bb}$ for the 4$b$ BG, we present the following distributions in ~\ref{fig:bg-dist} below:
a) from the left frame, one can see that for the steeply falling 
$M_{4b}$ distribution, increasing the $M_{4b}$ cut from 1 TeV to 1.5 TeV would reduce this BG by about one order of magnitude;
b) from the right frame, one can see that decreasing the upper cut on
$\Delta R_{bb}$ (for one of the pairs chosen according to the procedure described above) would also significantly reduce the BG.
\begin{figure}[htb]
\hspace*{-0.3cm}\includegraphics[width=0.57\textwidth]{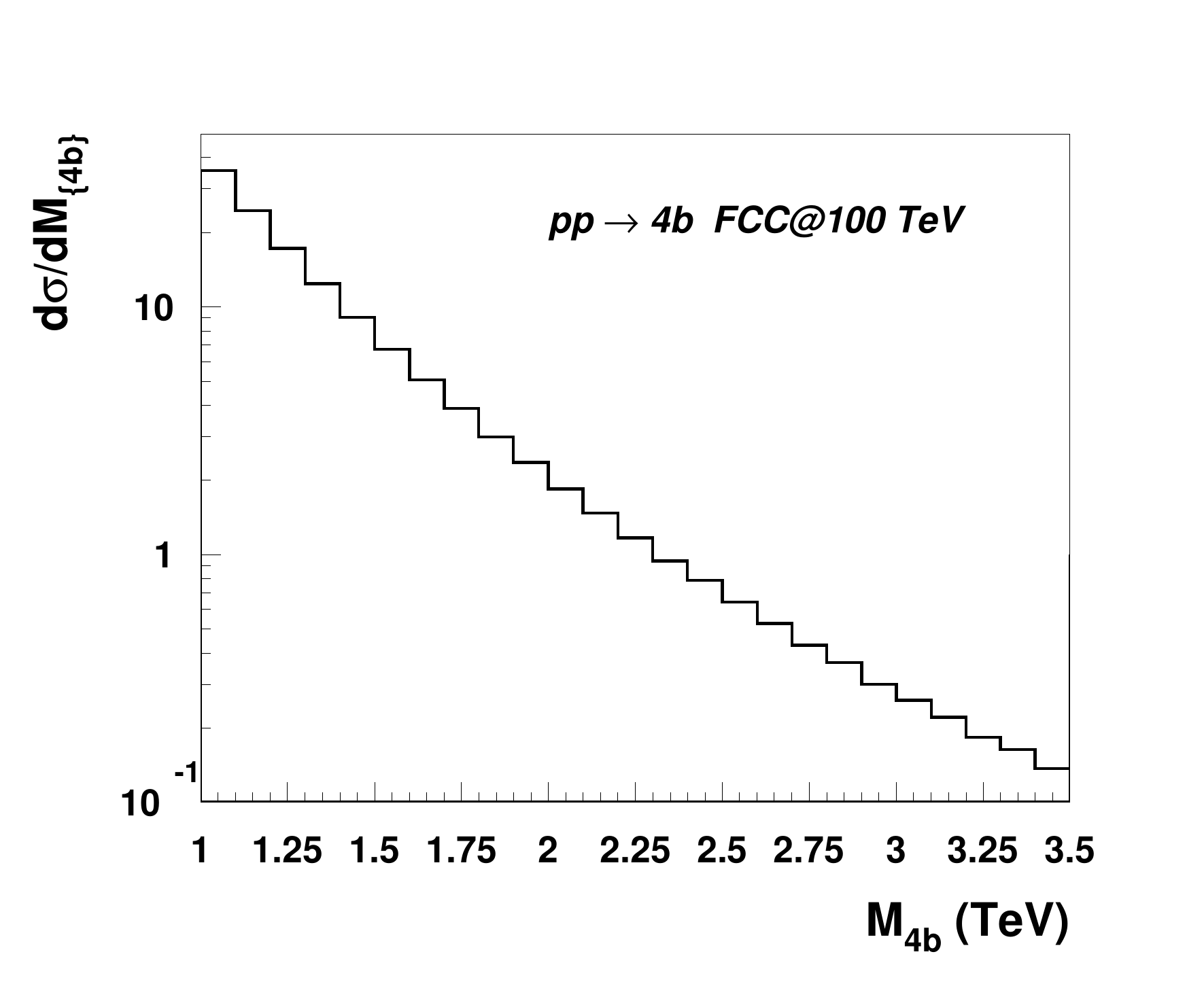}%
\hspace*{-1cm}\includegraphics[width=0.57\textwidth]{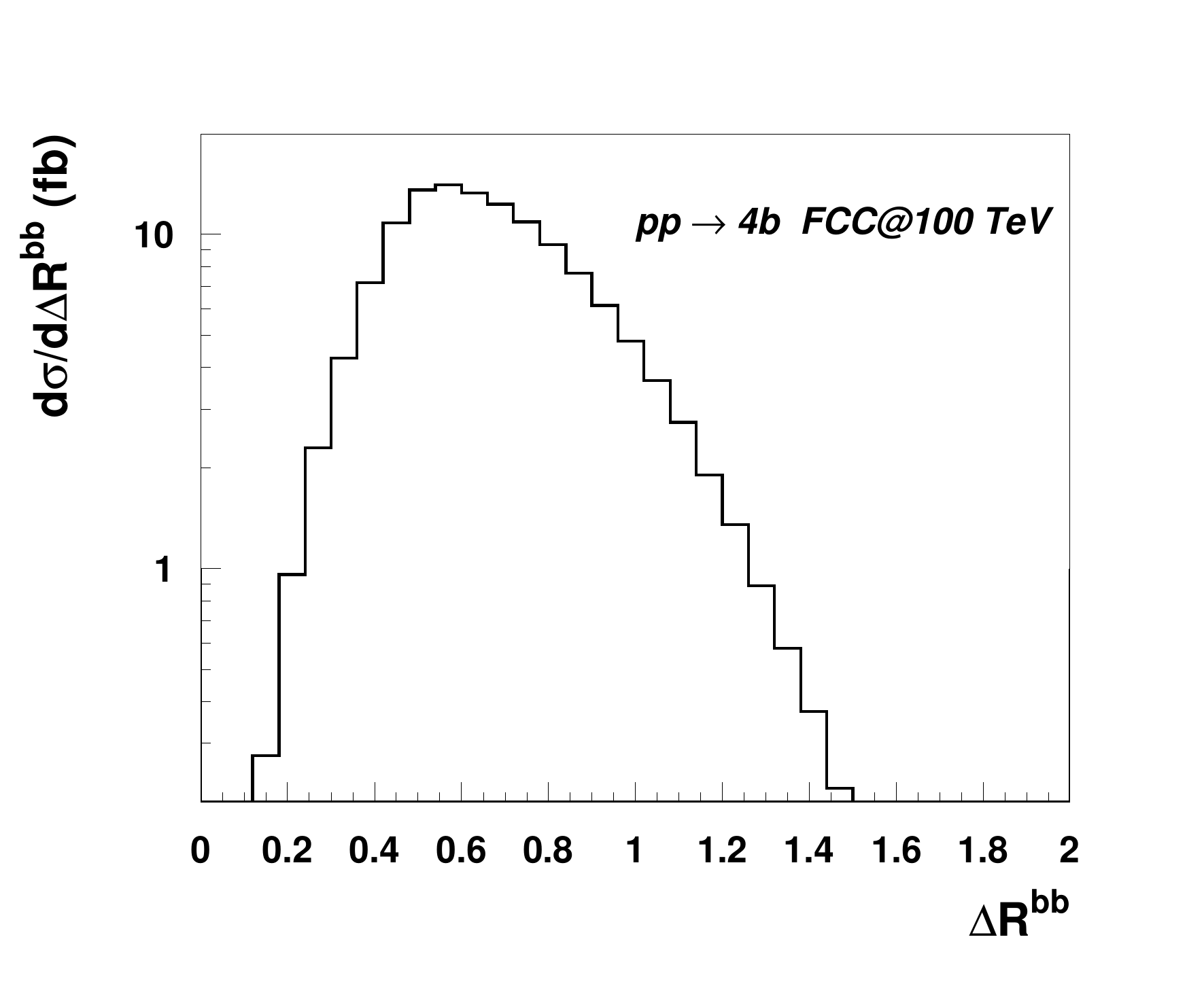}
    \caption{Distributions of $M_{4b}$ (left) and $\Delta R_{bb}$ (right) for the process
    $pp\to b\bar{b}b\bar{b}$ used for the 6$b$ BG generation via \texttt{PYTHIA}
    for the cuts \eqref{eq:mhcut} -- \eqref{eq:etacut} applied at parton level. }
    \label{fig:bg-dist}
\end{figure}
When applying the cuts \eqref{eq:mhcut}--\eqref{eq:drcut}, using \texttt{CTEQ6l1} as PDF and setting the QCD scale equal to $M_{4b}$, the cross section (which we then use for the 6$b$ BG estimation) is found to be equal to 
19.0 fb.
As described above, we have used 4$b$ BG events to find the probability $\omega_{bb}$ to create an additional $b\bar{b}$ pair with $|M_{bb}-M_{h}|  \leq \Delta^{\text{cut}}_{M_h}=15~{\rm GeV}$ for various values of the $\Delta R^{\text{cut}}_{bb}$ cut.
After running 500K events through \texttt{PYTHIA}, the respective error for $\omega_{bb}$ lies at the percent level.
The results are presented in table~\ref{tab:DRweights} below and one can see right away that the
cut on $\Delta R_{bb}$ has the power to further reduce the SM BG.
For $\Delta R_{bb} <0.5$, $\omega_{bb}=8.6 \cdot 10^{-5}$ and $\sigma({\rm 6b})$ for the cuts \eqref{eq:mhcut}--\eqref{eq:drcut} can be estimated as:
\begin{align}
	\sigma({\rm 6b}) &= \sigma({\rm 4b}) \times \omega_{bb}(\Delta R_{bb} <0.5) \notag \\
	                 &= 19.0~{\rm fb} \times 8.6 \cdot 10^{-5} \notag \\
	                 &\simeq 1.6\cdot 10^{-3}~{\rm fb} \,.
\end{align}
\begin{table}[htb]
	\centering
	\begin{tabular}{r|cccc}
		\toprule
		$\Delta R_{bb} < \Delta R^{\text{cut}}_{bb}$ & 2.0                 & 1.5                 & 1.0                 & 0.5                 \\
		\midrule
		$\omega_{bb}$       & $1.1 \cdot 10^{-3}$ & $7.0 \cdot 10^{-4}$ & $3.5 \cdot 10^{-4}$ & $8.6 \cdot 10^{-5}$ \\
		\bottomrule
	\end{tabular}
	\caption{Probability $\omega_{bb}$ to create an additional $b\bar{b}$ pair from 4$b$ events with $|M_{bb}-M_{h}|  \leq \Delta^{\text{cut}}_{M_h}=15~{\rm GeV}$ for various values of $\Delta R^{\text{cut}}_{bb}$ cut as a result of running 500k 4$b$ events through \texttt{PYTHIA}.}
	\label{tab:DRweights}	
\end{table}

After the procedure of triple Higgs-jet tagging, the rate of the $hhh$ BG can be estimated as
\begin{equation}
	\sigma_{BG}({\rm hhh}) = \sigma(6b) \times \varepsilon_{hhh} \,,
	\simeq 9.5\cdot 10^{-5} fb
\end{equation}
while the signal rate is given by Eq.\ref{eq:sig}.

\begin{figure}[htb]
	\includegraphics[width=0.485\textwidth]{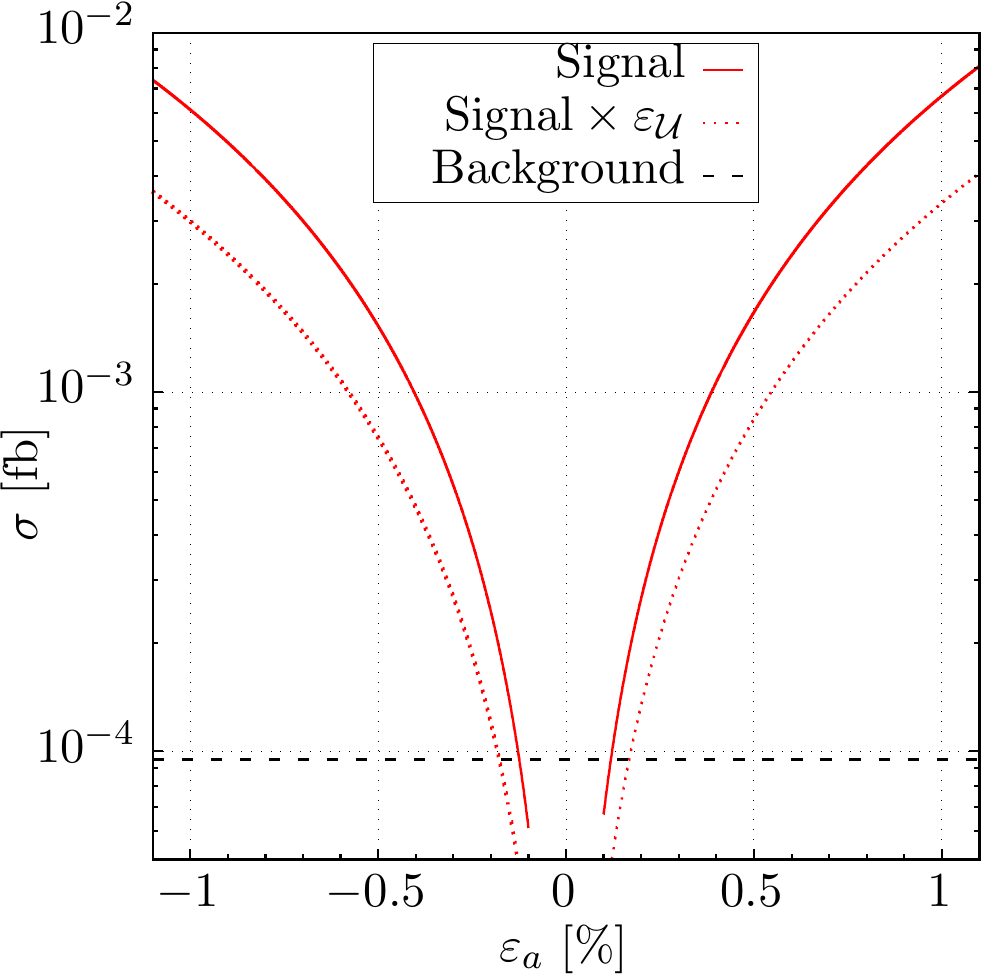}%
	\ 
	\includegraphics[width=0.515\textwidth]{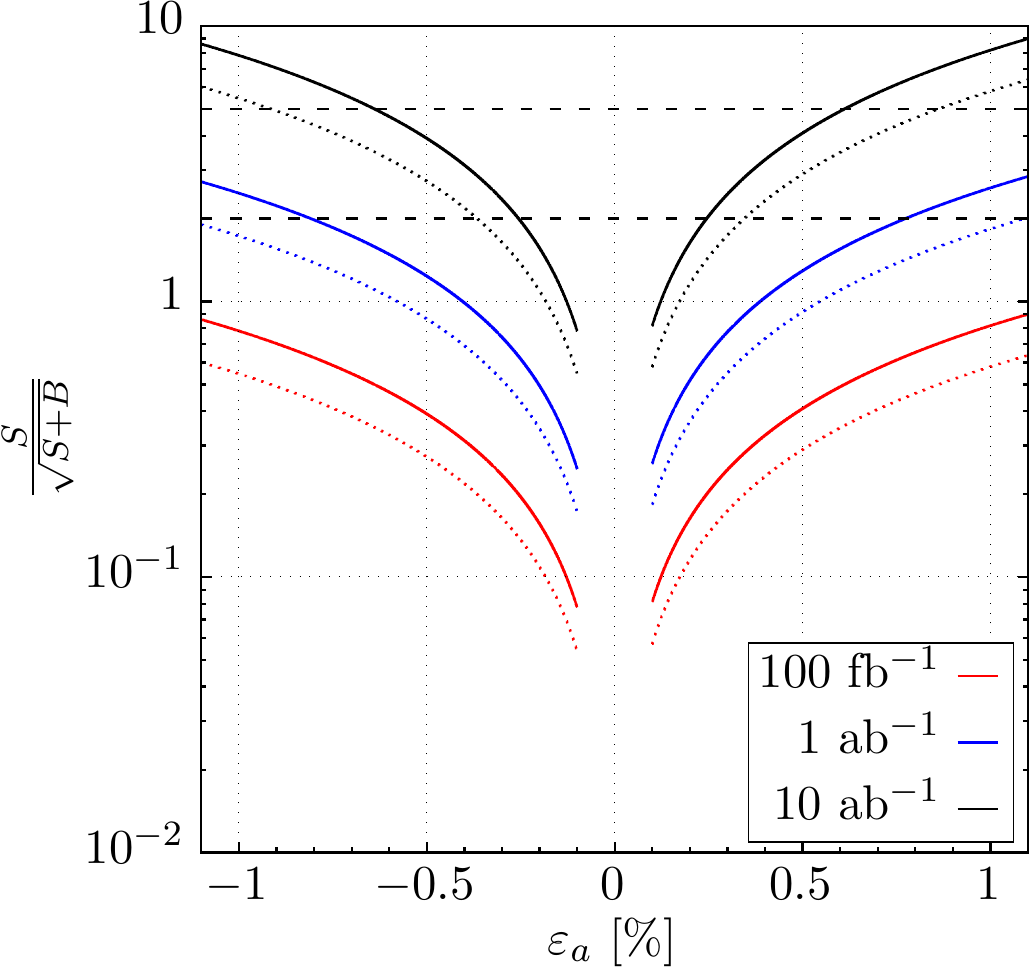}
    \caption{$\sigma_{\text{sig}}(hhh)$ and  $\sigma_{\text{BG}}(hhh)$
for $\varepsilon_a \in [-0.01 : 0.01]$ (left frame)
as well as the 100 TeV FCC sensitivity to $\varepsilon_a$ (right frame)
for 100 fb$^{-1}$, 1 ab$^{-1}$ 
and 10 ab$^{-1}$ integrated luminosities benchmarks. The dotted curves in both frames 
present results for the signal equal to $\sigma_{\text{sig}}(hhh) \times \varepsilon_\mathcal{U}$.
	}
    \label{fig:final-results}
\end{figure}
One can  check from equation~(16) of Ref.~\cite{Belyaev:2012bm}
that $\sigma(pp\to jj hhh)$ quite precisely scales as $\varepsilon_a^2=
(1-a)^2$, when $|\varepsilon_a| \ll 1$ and $\sigma(pp\to jjhhh) \gg \sigma(pp\to jjhhh)_{SM}$. Using this scaling and the rates from
table~\ref{tab:Mhhh} 
one can easily find the signal rates for smaller values of $\varepsilon_a$.
In Fig.~\ref{fig:final-results} we present, 
$\sigma_{sig}(hhh)$ and $\sigma_{\rm{BG}}(hhh)$
for $\varepsilon_a$ in the range $[-0.01 : 0.01]$ (left frame)
as well as the 100TeV FCC sensitivity to $\varepsilon_a$ (right frame).
One can see that the signal dominates over the 6$b$ BG and
becomes comparable to the 6$b$ BG only for $|\varepsilon_a|$ at the permille level or below.
The dotted curves in both frames present results for the signal equal to $\sigma_{\text{sig}}(hhh) \times \varepsilon_\mathcal{U}$ to take into account the cut  of the region of the parameter space where 
unitarity is violated.

One should note, that our BG estimation should be considered as an 
{\it upper bound} for the BG, since after the requirement of two additional
forward-backward jets, the actual BG is expected to be two orders of magnitude below just the 
6$b$ BG. Therefore, we can safely assume that for $|\varepsilon_a|>10^{-3}$,
the actual BG is negligible in comparison to the signal, hence it is only a question of luminosity to probe $\varepsilon_a$ up to the permille level.
For example, with 100 fb$^{-1}$, 1 ab$^{-1}$ 
and 10 ab$^{-1}$, one can probe $|\varepsilon_a|\simeq 2.5\cdot 10^{-2}$, 
$|\varepsilon_a|\simeq 7.5\cdot 10^{-3}$ and  $|\varepsilon_a|\simeq 2.5\cdot 10^{-3}$ respectively. We have used two standard deviations
criteria to judge about this sensitivity, which is indicated in the right pane of Fig.~\ref{fig:final-results} together with the 5$\,\sigma$ discovery limit in form of two horizontal lines at 2 and 5 respectively. Altogether, one can see that with triple Higgs VBF signatures at a 100 TeV FCC, we will be able to measure the $hVV$ coupling with permille accuracy.
This accuracy is remarkable since it is about  two orders of magnitude better than the sensitivity achievable  at  the LHC.

\section{Conclusions}
\label{sec:conclusion}
We have explored the potential of future hadron colliders
to test the couplings of a Higgs boson to gauge bosons.
As has been shown previously, if the coupling of the Higgs boson to gauge bosons deviates from the Standard Model, multi-boson production via vector-boson scattering 
can be hugely enhanced in comparison to the SM due to the lack of cancellation 
in longitudinal vector boson scattering.
Among these processes, triple Higgs boson production plays a special role ---
its enhancement is especially spectacular due to the absence of background from transversely polarised 
vector bosons in the final state.
While the rates from $pp\to jjhhh$ production in vector boson fusion are too low at the LHC and even 
at future 33 TeV $pp$ colliders, we have found that the 100 TeV $pp$ FCC has the unique opportunity to probe the $hVV$ coupling far beyond the LHC sensitivity using triple Higgs production via vector boson fusion.

We have evaluated the $pp\to jjhhh$ rates as a function of the deviation from the $hVV$ coupling, $\varepsilon_a$,
before and after VBF cuts and have estimated the 6$b$-jet background --- which turns out to be 
much smaller than the signal for $|\varepsilon_a|>10^{-3}$ ---
and have found that the 100 TeV $pp$ FCC can  probe this coupling with high precision.
A summary of our findings is presented in Fig.~\ref{fig:final-results},
demonstrating the impressive sensitivity to the $hVV$ coupling of the 100 TeV $pp$ FCC
via $hhh$ production in vector boson fusion up to permille accuracy. This sensitivity, which is about two orders of magnitude 
better than the sensitivity reachable at the LHC,
highlights a special role of the $hhh$ VBF production and stresses once more
the importance of the 100 TeV $pp$ FCC.

%\appendix
%\section{Some title}
%Please always give a title also for appendices.

\section*{Acknowledgements}

The authors acknowledge the use of the IRIDIS High Performance Computing Facility, and
associated support services at the University of Southampton, in the completion of this work.
AB would like to thank Douglas Ross for valuable discussions on the  estimation of  the QCD background.
AB and  PBS also grateful to  Micheleangelo Mangano for various discussions and help with ALPGEN package, 
Alexandra Carvalho for discussion and help with understanding efficiencies for Higgs-jet tagging.
AB and PBS acknowledge partial support from the InvisiblesPlus RISE from the European
Union Horizon 2020 research and innovation programme under the Marie Sklodowska-Curie grant
agreement No 690575. 
AB acknowledges partial  support from the STFC grant ST/L000296/1.
AB also thanks the NExT Institute, Royal Society Leverhulme Trust Senior Research Fellowship LT140094, Royal Society Internationl Exchange grant IE150682 and
Soton-FAPESP grant.
AB also acknowledge the support of IBS centre in Daejeon for the hospitality and support.

%\paragraph{Note added.} This is also a good position for notes added after the paper has been written.

\newpage

\bibliographystyle{bib}
\bibliography{bib}

\end{document}